\documentclass[twocolumn,english,aps,pra,reprint, superscriptaddress,showpacs,longbibliography, showkeys, nofootinbib]{revtex4-2}

\usepackage{amsmath,amssymb,bbm,mathrsfs,bm,braket,color,graphicx,comment,xcolor} 
\usepackage[colorlinks,citecolor=blue,urlcolor=blue,linkcolor = blue]{hyperref}
\usepackage[mathscr]{euscript}
\usepackage{enumitem}
\usepackage{dsfont}

\usepackage{float}
\makeatletter
\let\newfloat\newfloat@ltx
\makeatother
\usepackage{algorithm}
\usepackage[noend]{algpseudocode}

\begin{document}
\title{Low-depth Circuit Implementation of Parity Constraints for Quantum Optimization}

\author{Josua Unger}
\affiliation{Parity Quantum Computing GmbH, A-6020 Innsbruck, Austria}

\author{Anette Messinger}
\affiliation{Parity Quantum Computing GmbH, A-6020 Innsbruck, Austria}

\author{Benjamin E. Niehoff}
\affiliation{Parity Quantum Computing GmbH, A-6020 Innsbruck, Austria}

\author{Michael Fellner}
\affiliation{Parity Quantum Computing GmbH, A-6020 Innsbruck, Austria}
\affiliation{Institute for Theoretical Physics, University of Innsbruck, A-6020 Innsbruck, Austria}

\author{Wolfgang Lechner}
\affiliation{Parity Quantum Computing GmbH, A-6020 Innsbruck, Austria}
\affiliation{Institute for Theoretical Physics, University of Innsbruck, A-6020 Innsbruck, Austria}

\date{\today}

\begin{abstract}
We present a construction for circuits with low gate count and depth, implementing three- and four-body Pauli-Z product operators as they appear in the form of plaquette-shaped constraints in QAOA when using the parity mapping. The circuits can be implemented on any quantum device with nearest-neighbor connectivity on a square-lattice, using only one gate type and one orientation of two-qubit gates at a time. We find an upper bound for the circuit depth which is independent of the system size. The procedure is readily adjustable to hardware-specific restrictions, such as a minimum required spatial distance between simultaneously executed gates, or gates only being simultaneously executable within a subset of all the qubits, for example a single line. 
\end{abstract}

\maketitle

\section{Introduction}
\begin{figure*}
    \includegraphics[width=\textwidth]{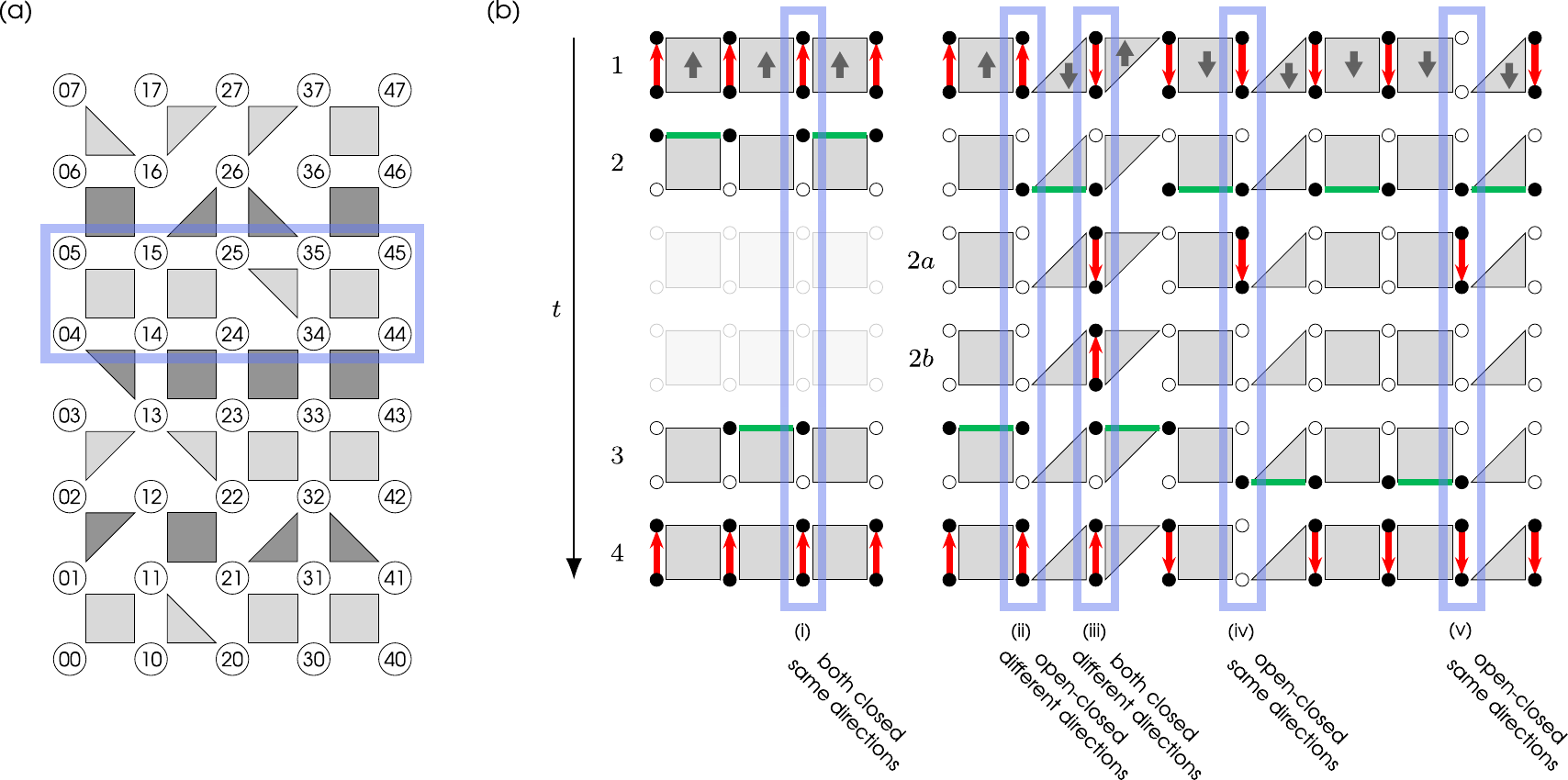}
    \caption{(a) Example layout of plaquette constraints (gray shapes) between qubits (empty, labelled circles) arranged on a square lattice. Qubit labels indicate the qubit coordinate. The implementation of the plaquette constraints is split into separate strips, of which one is highlighted with the blue box. The lighter and darker gray filling of the plaquettes indicates the two major steps needed for implementation, all strips of the same shade can be implemented in parallel. (b) Constraint implementation for a single strip with square plaquettes in four steps (left) or a sequence of square and triangle plaquettes in six steps (right). CNOT gates are represented by red arrows pointing from control to target, green solid lines lines represent ZZ gates. The gray arrows in the top row (time step $1$) indicate the CNOT gate direction chosen for each constraint in this example. Numbers on the left-hand side enumerate the time step of the drawn action. The blue boxes highlight examples of the different arrangements (direction of CNOT gates and open/closed at the boundary) which are possible between neighboring plaquettes and the corresponding adjustments to the gate sequence.}
    \label{fig:stripe_decomposition}
\end{figure*}

The field of quantum optimization has advanced rapidly in the last decades, due to an enormous amount of use cases from industry and academia that were revisited in the light of quantum optimization~\cite{lucas2014ising, yarkoni2022quantum, orus2019quantum, bauer2020quantum} and because of the research process on tackling the difficulties quantum computers are still facing. Addressing one of the main issues of current quantum devices, which is the connectivity problem, the Lechner-Hauke-Zoller (LHZ) architecture was introduced in 2015~\cite{Lechner2015}. It allows one to reformulate arbitrary optimization problems using single-body terms by introducing so-called parity constraints, which can be implemented with local three- and four-body interactions for quantum annealing~\cite{farhi2001,johnson2011, Albash2018} and for the quantum approximate optimization algorithm (QAOA)~\cite{Farhi2014, Lechner2020}. A generalization of the LHZ architecture to hyper-graphs, which we call the Parity Architecture~\cite{Ender2023}, was recently shown to have a significant advantage in the number of two-qubit gates for artificially constructed problem instances as well as for toy models for real-world use cases~\cite{Fellner2023}. Recent research has generalized the QAOA~\cite{Hadfield2019} and investigated phenomena like parameter concentration~\cite{Streif2020, Akshay2021, Wurtz2021}, confirming the QAOA as a promising candidate to show quantum advantage within the noisy intermediate scale quantum (NISQ) era~\cite{Preskill2018}. While in principle, any structure and arrangement of parity constraints is possible and implementations of the resulting constraint operations are known (see for example \cite{Holmes2020, NielsenChuang2011, Cowtan2020}), a particularly promising choice is a mapping such that all constraints are arranged as square and triangular plaquettes on a square lattice~\cite{Ender2023} (see for example the plaquettes drawn in Fig.~\ref{fig:stripe_decomposition}a). For some specific cases of such constraint arrangements, parallelizable QAOA circuits have been proposed \cite{Lechner2020, Sriluckshmy2023}.

In this work we derive a low-depth circuit to implement the operators corresponding to any such constraint arrangement using the QAOA. We exploit the structure of the constraint arrangement to minimize the circuit depth as well as the number of two-qubit gates required in the circuit. The proposed circuit construction leads to an improvement of the parity-mapped implementation of optimization problems on fully connected graphs shown in Ref.~\cite{Lechner2020}, but also represents a generalization of the circuit parallelization to arbitrary graphs and hyper-graphs, for which, to the best of our knowledge, no efficient implementation strategy has been presented yet. 

We furthermore show that the methods can be readily modified to accommodate restrictions of quantum hardware concerning the maximal possible gate parallelization (ratio of gate count to circuit depth). For platforms in which gates can be parallelized only along single lines of qubits, or in which gates can only be performed in parallel if the involved qubits have a certain physical distance, the resulting circuit is close to optimal in that it makes use of almost all parallelization realizable on the hardware. A specific example of such a restriction are hardware platforms based on Rydberg atoms where the Rydberg blockade prohibits the simultaneous execution of multi-qubit gates in a certain radius (cf. Ref.~\cite{Saffman2010, Henriet2020, Cong2022} and section~\ref{sec:parall_restrictions}).

The remainder of this paper is organized as follows. In Sec.~\ref{sec:Constraint_Implementation}, we review the implementation of parity constraints for QAOA, focusing on different decompositions of the respective operators into two-qubit gates. We then introduce the core work, the construction of the optimized circuit implementation for a given configuration of plaquette constraints, in Sec.~\ref{sec:Fully_parallel} and adapt it to hardware restrictions in Sec.~\ref{sec:parall_restrictions}. In Sec.~\ref{sec:performance_analysis}, we finally discuss how the resulting circuit depth and gate counts depend on system size, constraint configuration and hardware restrictions.

\section{Implementation of Parity constraint operators}\label{sec:Constraint_Implementation}
An optimization problem encoded in a logical Ising Hamiltonian~\cite{lucas2014ising} of the form
\begin{equation}
    H_p=\sum_{i=1} \tilde J_i \tilde Z_i + \sum_{i, j} \tilde J_{ij} \tilde Z_i \tilde Z_j +\sum_{i, j, k}\tilde  J_{ijk} \tilde Z_i\tilde Z_j \tilde Z_k + \dots
\end{equation}
is usually difficult to implement on a quantum device, as it requires multi-qubit and long-range interactions. Here, $Z_i$ denotes the Pauli $Z$-operator acting on qubit $i$ and the tilde indicates that the operators are logical. The Parity mapping maps each interaction term to a single physical qubit, e.g., ${J_{ijk}\tilde Z_i\tilde Z_j \tilde Z_k\mapsto J_m Z_m}$, thereby enlarging the Hilbert space from $N$ to ${K>N}$ qubits.
That allows one to get rid of the tedious long-range and multi-qubit interaction, however at the cost of an enlarged Hilbert space that has to be restricted with ${K-N}$ constraints. These constraints can be chosen to be (short-range) three- or four-body products that stabilize the logical subspace of the physical Hilbert space, i.e., the physical states that correspond to a logical state~\cite{Lechner2015}. 

\begin{figure}
\includegraphics[width=1\columnwidth]{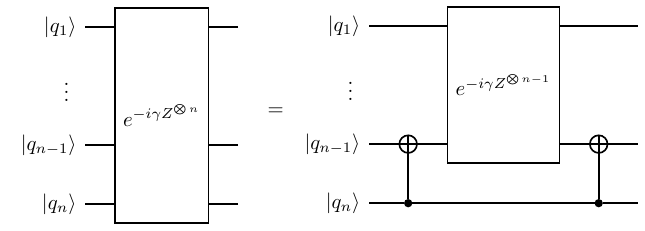}

\caption{Decomposition of constraint operators into CNOT gates and lower-order Z interactions. Recursively applying this decomposition yields a single-body $R_z$ rotation with two sequences of CNOT gates around it.
\label{fig:ZZ-to-CNOT}}

\end{figure}

These three [four]-body  constraints arising in the parity mapping are of the form
\begin{equation}
    Z_i Z_j Z_k [Z_l] \ket{\psi} = \ket{\psi}
\end{equation}
and are usually enforced via an energy penalty in the problem Hamiltonian. The mapped Hamiltonian is then
\begin{equation}
    H_\text{parity} = \sum_{i=1}^K J_i Z_i + c\sum_{l=1}^{K-N}C_l
\end{equation}
with ${C_l = Z_{l_1}Z_{l_2}Z_{l_3}[Z_{l_4}]}$,
where $c$ denotes the penalty strength for violated constraints and the brackets indicate that the fourth qubit does not occur in all constraints.
The constraints can always be chosen~\cite{terHoeven2023} such that the resulting QAOA operators for the constraint term are then of the form
\begin{equation}
    e^{i \alpha Z_i Z_j Z_k [Z_l]},
\end{equation}
where the Pauli operators $Z_i$, $Z_j$, $Z_k$ and $Z_l$ act on four qubits forming a unit cell of a square lattice. In the following, we show how any such plaquette constraint can be implemented with a sequence of controlled-NOT (CNOT) and ZZ gates, with
\begin{equation}
\mathrm{CNOT}_{jk}=|0\rangle\langle0|_{j}\mathds{1}_{k}+|1\rangle\langle1|_{j}X_{k}
\end{equation}
and
\begin{equation}
\mathrm{ZZ}_{jk}(\alpha)=e^{\mathrm{i}\alpha Z_{j}Z_{k}}.
\end{equation}
An operator $e^{i\alpha Z^{\otimes n}}$ can be decomposed into an operator $e^{i\alpha Z^{\otimes n-1}}$ and CNOT gates as depicted in Fig.~\ref{fig:ZZ-to-CNOT}. For this, the CNOT gates must be controlled by the $n$-th qubit and target (any) one of the remaining qubits.
For operators of arbitrarily high order, this method can be iteratively applied until only a two-body term, implementable
with a ZZ gate, or even a single-body term, implementable with a single-body Z rotation, remains.

This procedure allows us, for example, to implement a three-body constraint operator
with the gate sequence 
\begin{equation}\label{eq:3body_decomposition}
e^{i\alpha Z_{A}Z_{B}Z_{C}} =\mathrm{CNOT}_{AB}\mathrm{ZZ}_{BC}(\alpha)\mathrm{CNOT}_{AB}
\end{equation}
and a four-body operator with
\begin{equation}\label{eq:4body_decomposition}
\begin{split}
e&^{i \alpha Z_{A}Z_{B}Z_{C}Z_{D}} =\\
&\mathrm{CNOT}_{AB}\mathrm{CNOT}_{DC}\mathrm{ZZ}_{BC}(\alpha)\mathrm{CNOT}_{DC}\mathrm{CNOT}_{AB}.
\end{split}
\end{equation}
This ensures that any three- or four-body plaquette constraint (i.e., a constraint having all its qubits in a single unit cell of the lattice) can be implemented with nearest-neighbor interactions and in circuit depth three, as the two initial and the two final CNOT gates required for the four-body constraint [see Eq.~\eqref{eq:4body_decomposition}] can be implemented in parallel, respectively. Note that a variety of other decompositions is possible.

A decomposition as in Eq.~\eqref{eq:3body_decomposition} also allows for the implementation of multiple constraints at the same time. For example, the product of two operators $e^{i\alpha Z_{A}Z_{B}Z_{C}}$
and $e^{i\beta Z_{A}Z_{B}Z_{D}}$ can be implemented using only a single CNOT gate in the beginning and the end as
\begin{equation}\label{eq:constraint_product}
\begin{split}
e^{i\alpha Z_{A}Z_{B}Z_{C}}&e^{i\beta Z_{A}Z_{B}Z_{D}}=\\
&\mathrm{CNOT}_{AB}\mathrm{ZZ}_{BC}(\alpha)\mathrm{ZZ}_{BD}(\beta)\mathrm{CNOT}_{AB}.
\end{split}
\end{equation}
For comparison, decomposing each constraint separately would require twice as many CNOT gates. When putting the two separate constraint circuits together, one can also see that two CNOT gates (one in the end of the first circuit, and one in the beginning of the second) cancel each other, which leaves us with the same result.


This example illustrates how any two plaquette constraints which share two qubits at their boundary can in principle be implemented with fewer gates and a smaller circuit depth by `sharing' the same CNOT gate at the boundary.

Based on these observations, we derive a parallelized implementation of arbitrary plaquette layouts.
Note that, while we describe all circuits in terms of CNOT and ZZ gates, they can be easily translated to other universal gate sets. A ZZ gate, for example, can just be decomposed one step further with the procedure described in Fig.~\ref{fig:ZZ-to-CNOT}. Alternatively, a CNOT gate can be transformed into a single ZZ gate by adding single-body operations.

\section{Fully parallel implementation}\label{sec:Fully_parallel}
Let us consider a rectangular ${n\times m}$ grid of qubits with positions ${(i, j)}$.
We start by slicing the layout of plaquette constraints into horizontal strips, where each strip represents a single row of plaquettes. A single strip contains the qubits at positions ${\{(i, j)|1\leq i \leq n\}}$ and ${\{(i, j+1)|1 \leq i \leq n\}}$ for a fixed vertical position $j$, as for example the highlighted region in Fig.~\ref{fig:stripe_decomposition}a. 
Note that slicing the layout vertically works analogously and can lead to a different final circuit depth. Therefore, in order to obtain an optimal result, both versions should be considered for optimization and the one with smaller circuit depth used in the end. For demonstrative purposes, we focus only on the horizontal slicing in this work.

We determine a circuit to implement the plaquettes on each strip separately, and in the end combine them to a final circuit. As every strip only shares qubits with the two adjacent strips, the circuits of every second strip can be implemented in parallel. It is noteworthy that the algorithm outlined here is deterministic for a given layout.

In what follows, we describe the different plaquette configurations that can occur within a single strip at the vertical positions $j$ and ${j+1}$ and how the corresponding constraints are implemented.

\subsection{Trivial case: Square plaquettes only}\label{subsec:trivial_case_squares}
In a strip containing only square plaquettes, every neighboring pair of plaquettes shares both of the qubits
on the boundary. In order to exploit Eq.~\eqref{eq:constraint_product}, we therefore decompose the constraint operator using CNOT gates along exactly these edges, as shown in Fig.~\ref{fig:stripe_decomposition}b (left side).
The entire constraint circuit for this strip can then be implemented by the following procedure:
\begin{description}
\item[Step 1] Apply vertical CNOT gates along all edges between constraints and on the sides of the strips, controlling the bottom and targeting the top qubit,
\begin{equation}
\prod_{i=1}^n \text{CNOT}_{(i,j), (i,j+1)}.
\end{equation}
\item[Steps 2\&3] Apply horizontal ZZ gates between the two top qubits of every constraint, corresponding to
\begin{equation}
\prod_{i=1}^{n-1} \text{ZZ}_{(i,j+1), (i+1,j+1)}.
\end{equation}
\item[Step 4] Re-apply Step 1 and terminate.
\end{description}
While both, the gates of step 1 and of step 4, can be implemented in parallel, respectively, the middle part must be split into two steps as two gates cannot act on the same qubit at the same time.
Note that another possible choice is to reverse the direction of all CNOT gates and perform the ZZ gates on the qubits at the bottom of the respective constraints (i.e., swap $j$ and $j+1$ in the above instructions). In both cases, this implementation has a circuit depth of four. While this particular construction only implements sequences of four-body plaquettes, the described four steps form the basis of the circuit construction for all other scenarios.

\subsection{Arbitrary plaquette sequences}\label{subsec:mixed_plaquettes} 
In this section, we generalize the circuit for plaquette configurations containing triangle plaquettes, following the same structure. In particular, we decompose all constraints into a single horizontal ZZ gate surrounded by two or four vertical CNOT gates for three- or four-body constraints, respectively. As before, we exploit Eq.~\eqref{eq:constraint_product} and execute non-conflicting CNOT- and ZZ gates in parallel to reduce the gate count and the circuit depth.

Around triangle plaquettes (three-body constraints), the required gate sequence can vary. We can, however, implement any plaquette configuration by just adding or removing vertical CNOT gates at the boundary to such plaquettes as an adjustment to the procedure introduced in Sec.~\ref{subsec:trivial_case_squares}.

\begin{figure}
    \centering
    \includegraphics[width=\columnwidth]{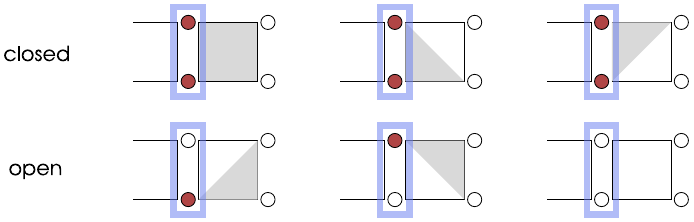}
    \caption{Illustration of plaquettes that are closed (upper row) or open (lower row) at the boundary on their left side (highlighted blue).  Note that the plaquette on the other side of the boundary does not have any influence on our definition of open and closed.}
    \label{fig:open_closed}
\end{figure}

Before we elaborate on that, let us introduce some terminology. We call a plaquette \textit{closed} at a boundary to another plaquette whenever both boundary qubits are included in the corresponding constraint. In any other case we call it \textit{open} (this means either that only one of the boundary qubits is included in the corresponding constraint, or that there is no constraint at the plaquette). This is also illustrated in Fig.~\ref{fig:open_closed}.

Only two pieces of knowledge at the boundary between two plaquettes are required to determine the necessary corrections to the circuit:
\begin{enumerate}
\item Whether the adjacent plaquettes are closed at the boundary or open. 
\item The direction of CNOT gates at each plaquette and, with that, the side of the strip on which the ZZ gate is performed.
\end{enumerate}

Note that, while the direction of CNOT gates can be freely chosen for square plaquettes, triangle plaquettes must be implemented with CNOT gates towards the side which includes two of the constraint qubits (in order to implement the ZZ gate there). Plaquette positions at which there is no constraint (in the beginning and end of the chip or if there are holes in the layout) are not associated with a particular CNOT gate direction. We will interpret them as having whichever direction results in the least corrections at each boundary (in fact, if we do this, there are never any additional gates necessary at such boundaries).
The goal is now to choose the direction of CNOT gates in sequences of square plaquettes (\textit{regions}), such that the necessary adjustments at boundaries to triangle plaquettes result in the smallest possible circuit depth and gate count. For this, we look at the different cases and adjustments that can become necessary in the following.  Note that for fixed CNOT gate directions, all adjustments are completely local to the boundaries between plaquettes, i.e., in order to determine the adjustments at a certain boundary, no knowledge about the boundaries on the other sides of the adjacent plaquettes is required.

The different possible boundary cases (i)-(v) are illustrated on an example strip in Fig.~\ref{fig:stripe_decomposition}b, their required circuit adjustments are explained in the following. The different cases may cause an additional circuit depth in the implementation. Therefore, we associate each case with a cost $c_\text{d}$ corresponding to its depth overhead.

\subsubsection*{Both plaquettes closed, same direction}
Whenever both adjacent plaquettes are closed at the boundary of interest and both have the same direction of CNOT gates as in case (i) in Fig.~\ref{fig:stripe_decomposition}b, no additional adjustment is needed at the boundary position. The plaquettes can be square or triangle plaquettes, as long as they are closed at the boundary. Cost: ${c_\text{d}=0}$.

\subsubsection*{Both plaquettes closed, different directions}
If the two constraints are implemented with CNOT gates of opposing directions [see case (iii) in Fig.~\ref{fig:stripe_decomposition}b] and thus with ZZ gates on different sides, we cannot use the same CNOT gate for both constraint decompositions. The direction of the first CNOT gate at the boundary should match the CNOT gate direction of the plaquette whose ZZ gate is implemented first (in step 2). After that ZZ gate, the initial CNOT gate must be repeated to complete the implementation of the first constraint. Only then can one apply another CNOT gate of opposite direction before the second ZZ gate (step 3). The final CNOT gate must then target the side of the later ZZ gate to complete the decomposition of that constraint as well. This implementation requires a circuit depth of $6$ and essentially corresponds to implementing the two adjacent constraints sequentially. Cost: ${c_\text{d}=2}$.

\subsubsection*{Closed plaquette and open plaquette, same direction}
If only one of the two plaquettes is closed at the boundary, but both constraints are decomposed such that their ZZ gate is on the same side [see cases (iv) and (v) in Fig.~\ref{fig:stripe_decomposition}b], we cannot use the trivial implementation either. The circuit would include the two boundary qubits in both constraints, but we want only one of the constraints to include them both. This means, only the ZZ gate of the closed constraint should be preceded and followed by CNOT gates at the boundary qubits, but not the ZZ gate of the open constraint.

If the ZZ gate corresponding to the closed constraint is applied before that of the open constraint [case (iv)], i.e., in step 2, we apply the CNOT gate in step 1 as usual, place an additional CNOT gate between the two ZZ gates (i.e., between steps 2 and 3), but leave out the final CNOT gate in step 4. Similarly, if the ZZ gate corresponding to the open constraint is applied first [case (v)], we leave out the initial CNOT gate (step 1) instead of the final one. The directions of the CNOT gates always must match to the implementation of the closed constraint (as this is the only constraint which requires CNOT gates here). This adjustment increases the circuit depth to $5$. Cost: ${c_\text{d}=1}$.

\subsubsection*{Closed plaquette and open plaquette, different directions}
As in the previous case, only one of the two constraints includes both qubits,  but the gate direction is not the same [see case (ii) in Fig.~\ref{fig:stripe_decomposition}b]. Since the two constraints are implemented with the ZZ gate at different sides now, the implementation is simple. In fact, it is the same as in the trivial case, with the CNOT gates at the boundary always matching the implementation of the closed constraint. This does not affect the open constraint, even though the two constraints
share a qubit: All CNOT gates at the shared qubit are controlled by that qubit and thus commute with the ZZ gate of the open constraint. Cost: ${c_\text{d}=0}$.

\subsubsection*{Both plaquettes open}
If both plaquettes are open at the boundary, no CNOT gates are required at this boundary at all. This is independent of whether the two constraints share a qubit or not, and does not lead to any increase in circuit depth. Cost: ${c_\text{d}=0}$.\\
\\
These five cases cover all situations and corresponding circuit adjustments which can occur within a strip. Note that additional CNOT gates due to such adjustments can be executed in parallel, so the total circuit depth is just the depth of the most expensive boundary configuration appearing in the strip, and thus has an upper bound of $6$. To minimize the circuit depth, the directions of CNOT gates for all square plaquettes should be chosen such that
\begin{itemize}
    \item neighboring square plaquettes always have the same direction and 
    \item the direction of a series of square plaquettes is such that the depth increase due to the configurations at the boundary to the next triangle plaquettes is minimal. If there is a triangle plaquette on both sides of a sequence of squares, choose their direction such that the maximum depth increase from both sides is the smallest.
\end{itemize}

Take for example the right block of the plaquette layout in Fig.~\ref{fig:stripe_decomposition}(b). Going from left to right, the first square plaquette borders a triangle plaquette only on one side (ii), so it's direction is chosen to minimize the depth of that boundary. The next square plaquette has two boundaries of the type "open-closed". With either choice of the plaquette direction, one of the boundaries would have "same direction" and one "different directions", so it does not matter in which direction we implement this plaquette.  Finally, we have two neighboring square plaquettes. Since they should be implemented in the same direction, we consider only two possible cases. If both are implemented upwards, the right boundary would be without depth increase but the left boundary would be "both closed - different directions" which corresponds to the maximum depth of 6. If both square plaquettes are implemented downwards as shown in the figure, the left boundary is without depth increase and the right boundary is "open-closed - same directions" which only requires a depth of 5. Therefore we should choose this direction. The overall procedure for determining the optimal gate sequence is summarized in Algorithm~\ref{alg:optimizer}.

\begin{algorithm*}
    \caption{Constraint Circuit Optimization}
    \label{alg:optimizer}
    \begin{algorithmic}[1]
        \State Divide the layout into stripes of plaquettes.\Comment{Do the division horizontally and vertically and use the shorter circuit.}
        \State $H \gets$ the set of horizontal stripes of plaquettes.
        \State $V \gets$ the set of vertical stripes of plaquettes.
        \For {$S\in \{H, V\}$}
            \For {$s\in S$} \Comment{Determine the circuit for all stripes.}
                \State $R \gets$  the set of all regions in the stripe.
                \Comment{A region is a sequence of 4-body plaquettes.}
                \For{$r\in R$}
                    \State $\bm{c} \gets [\,]$
                    \For{$\texttt{dir}\in \{\texttt{UP}, \texttt{DOWN}\}$}
                    \Comment{Try out both possible gate directions for the region.}
                        \State $\bm{c}[\texttt{dir}] \gets \texttt{cost}(r, \texttt{dir})$
                        \Comment{Calculate the associated additional depth as outlined in the main text.}
                    \EndFor
                    \State $r.\texttt{direction}\gets \min_\texttt{dir}(\bm{c})$
                    \Comment{Assign the direction yielding the minimal depth to the region.}
                \EndFor
                \State $s.\texttt{circuit} \gets$ \texttt{assembleStripeCircuit}($s, R$)
            \EndFor
            \State $\texttt{circuit}[S] \gets$ \texttt{assembleFullCircuit}($S$)
            \Comment{The circuits for each ${(d+1)}$-th strip can be executed in parallel.}
        \EndFor
        \State \textbf{return} $\arg\min(\texttt{depth}(\texttt{circuit}[H]), \texttt{depth}(\texttt{circuit}[V]))$
        \Comment{Finally, return the shortest of the two circuits.}
    \end{algorithmic}
\end{algorithm*}

\section{Implementation under parallelization restrictions}\label{sec:parall_restrictions}
\subsection{Minimal distance between simultaneous gates}
\begin{figure}
    \centering
    \includegraphics[width=0.85\columnwidth]{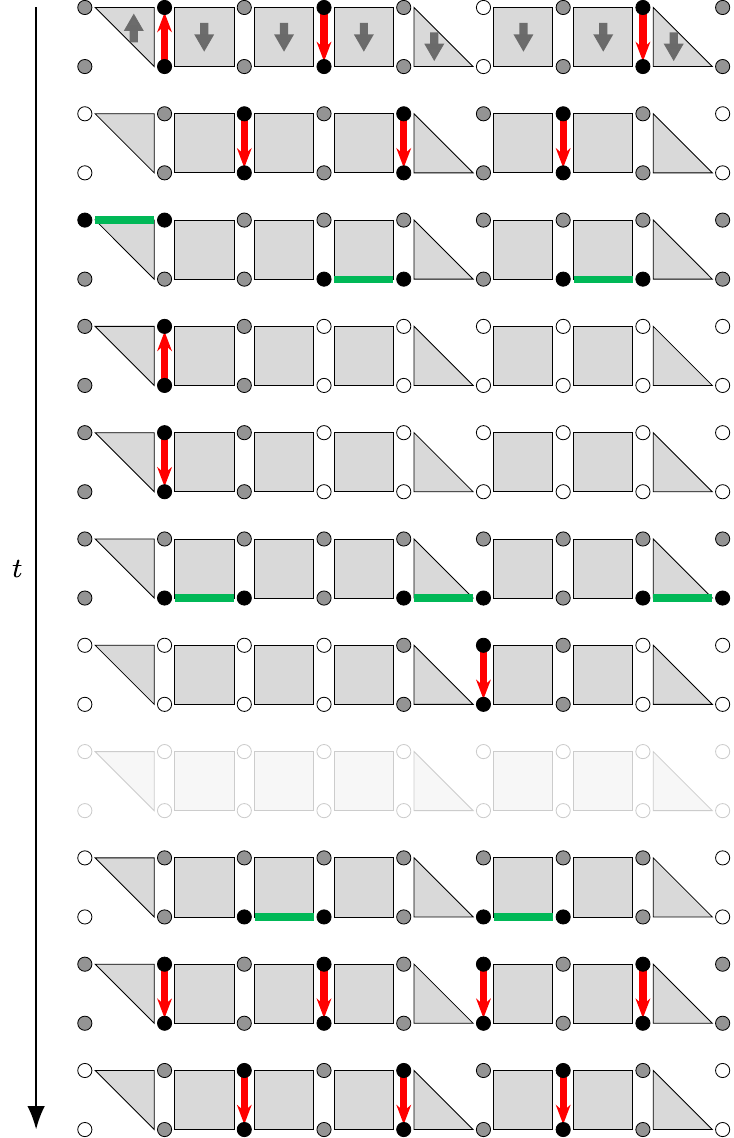}
    \caption{Schematic representation of the gate sequence to implement the constraints on an example strip where the minimal distance between simultaneous gates is ${d=2}$. Gates are represented in the same way as in Fig.~\ref{fig:stripe_decomposition}. In each step, gray qubits are blocked due to distance requirements to other gates in the vicinity and black qubits are actively involved in a gate. The shown example can be implemented in depth ${D=10}$. The worst case depth under the ${d=2}$ restriction is ${D=11}$, occurring if additional gates are required during the step faded out in the figure.}
    \label{fig:stripe_decomposition_restricted}
\end{figure}

The algorithm introduced above can be easily adjusted to run on hardware where neighboring or close-by gates can not be implemented in parallel, as for example when using Rydberg gates on atomic qubits\footnote{For atomic qubit platforms there exists an alternative proposal to implement constraints using four-body couplers~\cite{Dlaska2022}.}~\cite{Saffman2010, Henriet2020, Cong2022}. The typical restriction here is that two multi-qubit gates can only be performed in parallel if the minimal distance between any qubit involved in the first gate and any qubit involved in the second gate is larger than a certain constant $d$ (in units of the lattice constant), which we refer to as the parallelization distance. 
Choosing a higher integer ${d>1}$ can be beneficial in Rydberg devices since it makes simultaneously executed multi-qubit gates feel the potential of the respective other excited atoms less and thus cross-talk is reduced. In experimental realizations of Rydberg devices, $d$ typically takes values around $2-4$ \cite{bluvstein2022quantum}.
For ${d>1}$, we therefore make the following adjustments to the final circuit:

Within each strip, we split the gates of step 1 into $d$ consecutive moments in time such that every $d$-th gate is in the same moment. Instead of the next \textit{two} steps (steps 2 and 3), we split the horizontal gates into $d+1$ moments such that every ${(d+1)}$-th gate is in the same moment (note that every horizontal gate occupies two qubits along the strip so we need to split them into more moments than for the CNOT gates). Any additional vertical CNOT gates which were initially in-between steps 2 and 3 (labeled $2a$ and $2b$ in Fig.~\ref{fig:stripe_decomposition}b) must now be applied between the new moments at which the ZZ gates of the two adjacent constraints are implemented. Finally, step 4 is split into $d$ moments in the same manner as step 1. An example of such an implementation for $d=2$ is shown in Fig.~\ref{fig:stripe_decomposition_restricted}. Instead of two, there are now three steps with horizontal gates, and additional CNOT gates can occur between each of those steps.
Furthermore, instead of implementing every other strip in parallel, we now implement every ${(d+1)}$-th strip in parallel.

One can easily verify that this construction is in fact just a generalization of the fully parallel implementation, which then corresponds to the case ${d=1}$.

\subsection{Parallelization along lines of qubits}
\begin{figure}
    \centering
    \includegraphics[width=0.6\columnwidth]{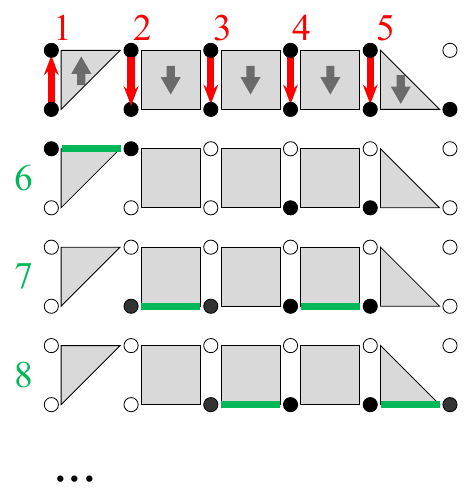}
    \caption{First few steps of a separation of a gate sequence into one-dimensional parallel slices. Vertical gates (e.g., steps 1-5) have to be implemented sequentially, but each step can be implemented together with corresponding steps in other strips. Horizontal gates (e.g., steps 6-8) have to be implemented for each strip separately, however, all gates in a single strip which are on the same side can be implemented in parallel.}
    \label{fig:1D_decomp}
\end{figure}

The presented circuit construction can also be used to create highly parallel circuits for platforms in which gates can only be applied in parallel if their qubits lie on the same one-dimensional line along the layout.
As each moment of the circuit contains either only horizontal or only vertical gates, we can easily split them into parallelizable sub-moments again, as depicted in Fig.~\ref{fig:1D_decomp}. Moments with horizontal gates are thus split into horizontal slices (each slice containing only gates from the same strip), and moments with vertical gates into vertical slices (containing up to one gate from every strip). 
This works for any parallel gate distance $d$ and still allows for a high degree of parallelization considering the given limitations.

\section{Performance analysis}\label{sec:performance_analysis}

\subsection{Circuit depth}\label{sec:circuit_depth}
In the following we investigate how large the depth $D$ of a circuit implementing an ${n \times n}$ square layout of qubits with three- and four-body constraints can get for large system dimensions $n$. All arguments and the resulting bounds in the following sections also apply to rectangular layouts in a straightforward way but we stick to square layouts for the simulations.

\begin{figure}
    \centering
    \includegraphics[width=\columnwidth]{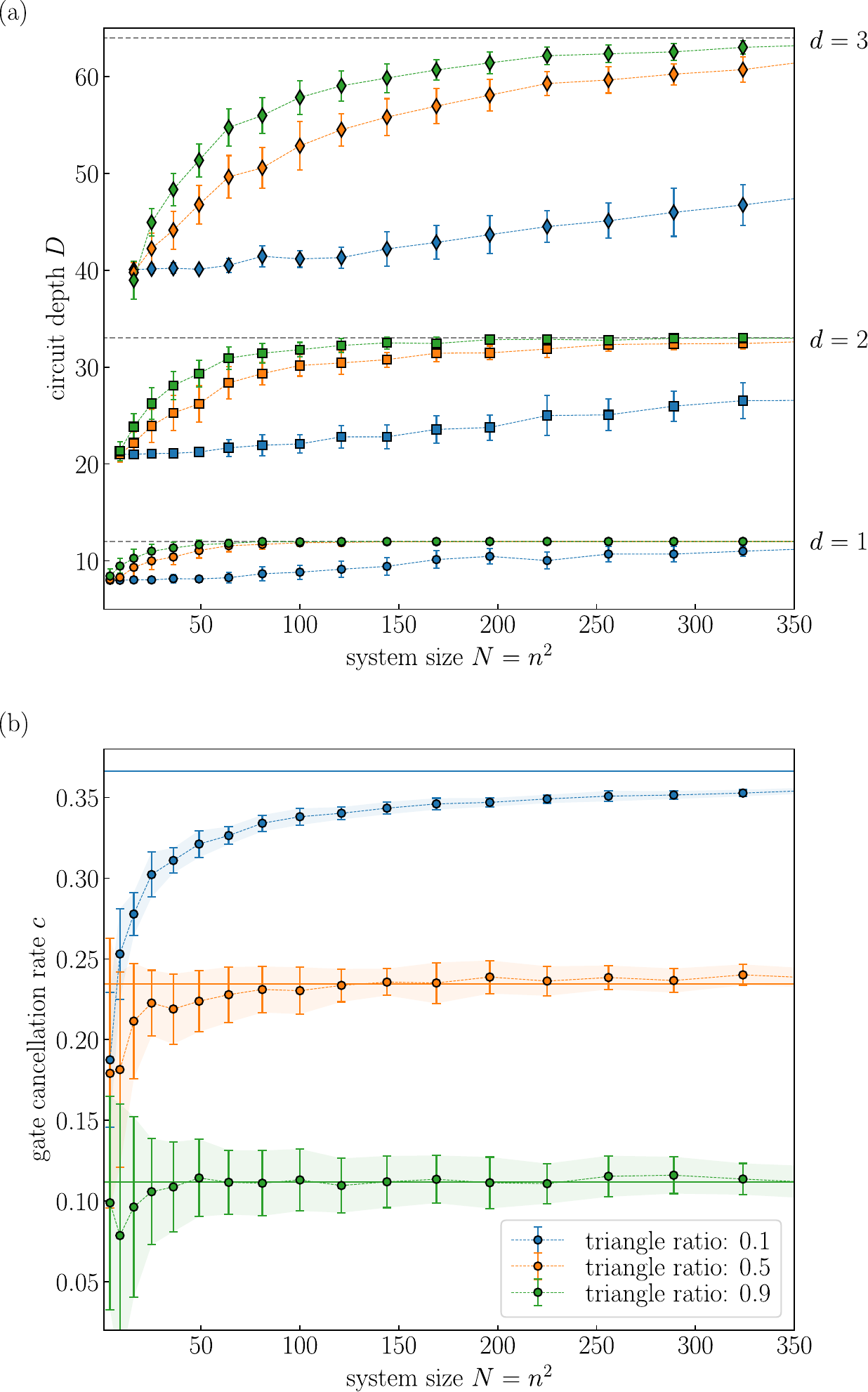}
    \caption{(a) Circuit depth for randomly sampled ${n \times n}$ layouts for various values of the triangle ratio $r_3$ and parallelization distances $d$. The gray dashed lines indicate the theoretical worst cases for a given parallelization distance. (b) Cancellation rates for randomly sampled ${n \times n}$ layouts for various triangle ratios. The solid horizontal lines represent the theoretical estimation for large system size. For both figures, the data points were averaged over 30 random layout and the error bars indicate their standard deviation.}
    \label{fig:performance_analysis}
\end{figure}

Recall from Fig.~\ref{fig:stripe_decomposition_restricted} that in one strip there are $2 d$ vertical CNOT gate moments in the beginning and end of the circuit and ${d+1}$ horizontal ZZ gate moments (here and in the following we are using the decomposition into CNOT and ZZ gates and count all two-qubit gates). Additionally, there are up to $2 d$ time slots for vertical CNOT gates between the ZZ gate steps. Even though they will not all be filled at each individual position in the strip, it can happen that each slot is occupied by a gate from somewhere in the strip or any other strips that is executed in parallel. Accounting for the number of strips, we multiply  with ${d+1}$  and obtain
\begin{equation}
    D \leq 5d^2 + 6d + 1
\end{equation}
as an upper bound on the depth.
Note that this is independent of the size of the chip or the number of constraints.
In Fig.~\ref{fig:performance_analysis}a we show how this worst case depth is approached for randomly chosen layouts for growing system size ${N=n^2}$ in dependence of parallelization distance and the ratio $r_{3}$ of three-body constraints relative to the total number of constraints. As expected, the depth grows up to a system size of ${(d+1)\times(d+1)}$ since there is no parallelization before that. Then, the depth approaches the worst case with a rate depending on the three-body constraint ratio due to the fact that the three-body constraints are fixed in their orientation and can cause conflicting orientations between constraints, requiring additional CNOT gates. Furthermore, it is apparent that for a larger parallelization distance the worst case is approached slower since there are more time slots in more strips that would need to be occupied in order to have a worst case depth.

For non-restricted hardware (${d=1}$), the circuit depth is ${D=12}$ in the worst case, and ${D=8}$ in the best case. This best-case scenario is reached, for example, for square-only layouts or layouts with only small numbers of triangle plaquettes which do not cause any conflicting situations. A well-known example of such a best-case scenario is the original LHZ layout~\cite{Lechner2015}, which encodes an all-to-all connected problem graph. So far, the best reported circuit depth for this constraint layout was ${28}$ for a decomposition into CNOT gates and Z rotations (${24}$ if we omit the depth increase from single-qubit gates)~\cite{Lechner2020}. For comparison, the corresponding depth of our implementation is ${D_\text{CNOT+Z}=12}$. At this point one should note that for specific cases, there can exist implementations with smaller circuit depth than that obtained with our procedure. For example, the implementation of a layout with exclusively square plaquettes (and with some adjustments also the LHZ layout) can be decomposed into CNOT gates and Z rotations in a depth of $D_\text{CNOT+Z}=10$ (not counting single-qubit gates), as shown in Appendix~\ref{appendix:lhzcase}. This specific implementation, however, exhibits a higher gate count and does not have an advantage for platforms with native ZZ gates.

\subsection{Gate count}
Similarly, we can determine the gate count $n_{\mathrm{g}}$ (the number of two-qubit gates necessary to implement the constraints) in the worst case situation. We start by looking at a single strip. In the case of three constraints there is a configuration of two three-body constraints with the same diagonal and their closed sides facing each other followed by a four-body constraint, 
\begin{center}
    \includegraphics[width=.35\columnwidth]{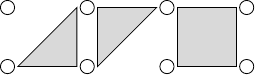},
\end{center}
which has an average gate count of $11/3$ per constraint and can be repeated with that average gate count. By exhaustive search and the freedom to determine the CNOT gate direction of four-body constraints on the boundaries of the strip, one can find only one configuration on three constraints that locally has a higher average gate count per constraint, a triangle in-between two squares where the direction of the square that shares a side with the triangle is opposite to the triangle direction. Extending this strip on both sides such that the CNOT gates of the squares are forced to point in the corresponding directions (which only has very few possibilities), we find that the average gate count per constraint drops below $11/3$ in all cases. Note that because at the boundary of a strip the directions are also not enforced, we can not have a strip that just consists of this pattern with higher average gate count.

For the other strips one can simply take the strip with the repeated $11/3$ pattern but displaced by one constraint relative to the previous strip. In doing so one finds that the vertical strips have the same pattern, therefore it is not possible to have a lower gate count by switching orientations. Thus the total gate count has the upper bound
\begin{equation}
    n_\text{g} \leq \frac{11}{3} N_C,
\end{equation}
where ${N_C=(n-1)^2}$ denotes the number of plaquettes in an ${n\times n}$ qubit layout. In a naive algorithm without optimisation the gate count could grow as $5N$ in the worst case. Note that the gate count does not depend on the parallelization distance.

We find that for randomly sampled constraint layouts the gate count exhibits a linear increase in system size $N$ with a rate that does not approach the worst case and slightly depends on the ratio $r_3$ of three-body constraints; for a higher ratio the increase is lower. Thus, even though there are more gate cancellations for longer sequences of four-body constraints, this is outweighed by the lower gate count of a single three-body constraint.

Furthermore, one can see from Fig.~\ref{fig:performance_analysis}b that the cancellation rate, defined as the number of gate cancellations normalized by the non-optimized gate count (if all constraints were to be implemented separately), approaches a constant $c$ depending on $r_{3}$ for large system sizes. For small layouts, there are also boundary effects since there can be no cancellation at a boundary but they drop off with roughly $1/n$, i.e., with the ratio of boundary to bulk gate positions. The asymptotic cancellation rate $c$ can be calculated to be 
\begin{equation}
    c = \frac{2[(1-r_{3}) + r_{3}/2][(1-r_{3}) + r_{3}/4]}{5(1- r_{3}) + 3r_{3}},
\end{equation}
where we neglected boundary effects and the fact that the orientation of the strips is chosen to minimize depth instead of gate count. To see why this formula holds asymptotically we start with a plaquette in a row. With probability $r_3$ it is a three-body constraint and from the four possible orientations only two can potentially lead to a cancellation with the next plaquette (those with a closed boundary on the right). In both these cases only one of the four possible three-body constraint orientations will give a cancellation of two gates and if a four-body constraint follows we can always orient it so that it leads to two gate cancellations. Thus we get a contribution of $2 r_3 \cdot 1/2 \cdot ((1-r_3) + r_3/4)$ to the expectation value of cancelled gates. Similarly, when starting with a four-body constraint in a given orientation we find that only one of the four three-body orientations leads to two cancelled gates and a four-body constraint can be oriented to lead to the cancellation.

Note that for a decomposition into CNOT and single-body Z-rotations, the depth increases only by $(d+1)^2$ multi-qubit gate steps (and another ${(d+1)^2}$ steps of exclusively single-body gates). This is independent of the arrangement or distribution of plaquettes, because in the decomposition presented before, the number of ZZ gates, and the time steps in which they are performed, is always fixed and only depends on the degree of parallelizability possible; all layout-dependent additional gates are already CNOT gates. For the special case of an ${n\times m}$ qubit layout with square plaquettes only, we arrive at a gate count of ${n_{g, \text{4-body}}=2m(n-1)}$, while for the original LHZ layout~\cite{Lechner2015, Lechner2020} with $n$ logical qubits we obtain ${n_{g, \text{LHZ}}=2(n-2)(n-3)}$.

\subsection{Comparison with other quantum circuit optimizers}
\begin{figure}
    \centering
    \includegraphics[width=\columnwidth]{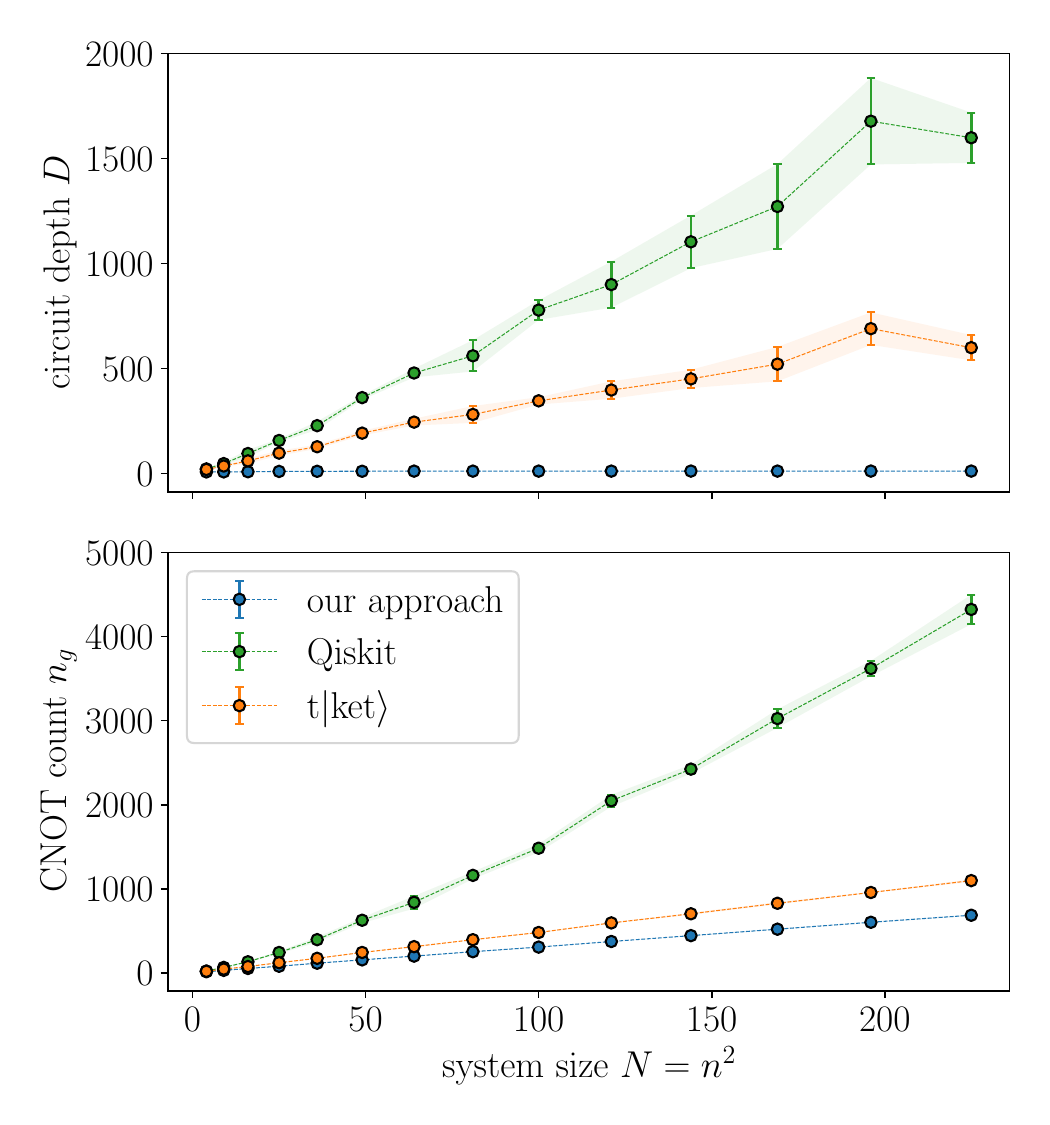}
    \caption{Comparison of circuit depth and gate count of the parity constraint implementation for different optimization approaches. Error bars indicate the standard deviation and are often smaller than the data points. Each data point represents the mean value of five random layouts with triangle ratio ${r_3=0.5}$.}
    \label{fig:optimizer_comparison}
\end{figure}

In order to set our results in context, we compare the gate count and the circuit depth obtained with our constructive optimization to numbers obtained with the heuristic circuit optimizers Qiskit~\cite{Qiskit} and $\text{t}|\text{ket}\rangle$~\cite{Sivarajah2021}. In particular for Qiskit we use the \texttt{transpile} method one a square grid with a gate set consisting of CNOT, Pauli rotations and Hadamard gates. For $\text{t}|\text{ket}\rangle$ we created a \texttt{Backend} with the same gate set and topology and \texttt{optimisation\_level} $=2$ and employed the \texttt{get\_compiled\_circuit} method. Figure~\ref{fig:optimizer_comparison} depicts the optimization results for circuits of different size for a triangle ratio ${r_3=0.5}$ and a parallel gate distance ${d=1}$, where the opimization starts after laying out the 3- and 4-body plaquettes on a square grid, i.e., the parity compilation is considered in the optimization process. For each data point, we averaged over five random layouts. As expected from the discussion in Sec.~\ref{sec:circuit_depth}, the circuit depth of our optimization approach converges at a comparably low number, while it increases with the system size for the other approaches. For our approach as well as for the $\text{t}|\text{ket}\rangle$ optimizer, we observe a linear scaling with $N$ (and therefore with $N_C$) for the CNOT count, while it scales roughly quadratically for the optimization with Qiskit. However, we note that this benchmark considers a very specific case of optimization for a particular implementation of constraints and can by no means be interpreted as a general meaningful benchmark of optimization techniques.

\section{Conclusion and Outlook}
We have presented a strategy to minimize the circuit depth and gate count to implement plaquette constraints on a square lattice layout. The strategy is applicable for arbitrary arrangements of three- and four-body constraints. For the original LHZ layout implementing fully connected graphs, the circuit depth and number of CNOT gates required is significantly smaller than in previous results~\cite{Lechner2020}.  The presented implementation is also useful for variations of QAOA in the parity architecture as for example the modular implementation introduced in Ref.~\cite{Ender2022} where the problem unitary contains only a sparse grid of plaquette constraints.

The regular arrangement of gates in the resulting circuit makes our strategy highly efficient for various possible hardware restrictions, and holds the potential for further benefits to specific hardware features. In particular, in the resulting circuit, every moment is always filled with only one type of gate. Simultaneous two-body gates are always oriented along the same axis. Furthermore, every strip along the layout is treated separately such that they can be implemented in any order or parallel grouping. Future work will have to extend this result to more arbitrary constraint shapes which are not restricted to plaquettes. 

An implementation of our approach on spin qubits was already investigated and simulation results suggest that it requires lower values for the gate fidelity than other QAOA implementation strategies~\cite{Ginzel2024}. In combination with suitable decoding strategies~\cite{Weidinger2023}, the presented optimized Parity QAOA circuit is therefore considered a promising candidate for implementing QAOA on spin qubits in a scalable way. Another promising lane for future research is to investigate the combination of our results with the insights from Ref.~\cite{Ender2023}, to further increase success probabilties for the quantum optimization.\\

\paragraph*{Acknowledgements -}
This work was supported by the Austrian Research Promotion Agency (FFG Project No. 884444, QFTE 2020), the Austrian Science Fund (FWF) through a START grant under Project No. Y1067-N27 and the SFB BeyondC Project No. F7108-N38.

\appendix
\renewcommand\thefigure{\thesection\arabic{figure}} 
\setcounter{figure}{0}    
\section{Alternative implementation for special case: Only square plaquettes}\label{appendix:lhzcase}
\begin{figure*}
    \centering
    \includegraphics[width=\textwidth]{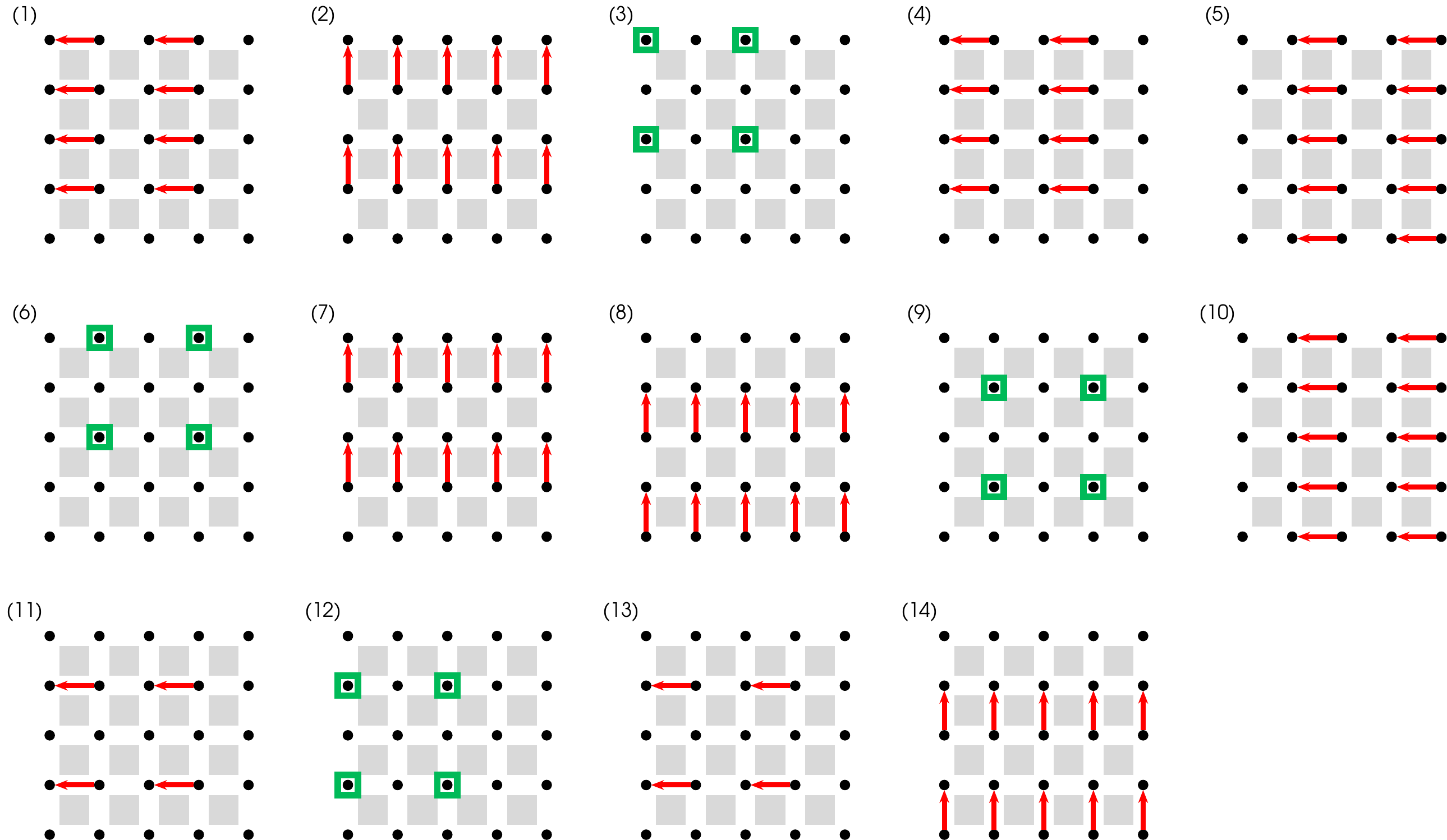}
    \caption{Gate sequence for implementing the QAOA circuit representing 4-body parity constraints on a ${5 \times 5}$ qubit lattice. The circuit depth does not increase for larger systems. CNOT gates are represented by red arrows (pointing from the control to the target qubit) and green squares depict single-qubit Z rotations. The total circuit depth is $14$, while the CNOT depth is $10$. Note that the circuit depth is larger when using ZZ gates.}
    \label{fig:special_case_depth}
\end{figure*}

For the decomposition into CNOT gates and Z rotations, it is possible to reduce the circuit depth to a total of $14$ steps if the plaquette layout only contains square plaquettes. Four of these steps only include single-qubit rotations; therefore the circuit shows a CNOT depth of ${10}$, which beats the optimal CNOT depth of $12$ from the procedure described in Sec.~\ref{sec:Fully_parallel} when decomposed to CNOT and Z gates. 
The gate sequence for a ${5\times 5}$ qubit layout is depicted in Fig.~\ref{fig:special_case_depth} and can be readily extended for larger systems, without affecting the circuit depth.

A drawback of this approach compared to the method described in the main text is that it requires a higher two-qubit gate count. The two-qubit gate count is 
\begin{equation}\label{eq:special_gate_count}
n_\text{g} = 4.5N_\text{C}+4\sqrt{N_\text{C}},
\end{equation}
compared to $n_\text{g}=4N_\text{C}+ \sqrt{N_\text{C}}$ for the approach discussed in the main text when decomposed to CNOT and Z gates. The terms proportional to $\sqrt{N_\text{C}}$ account for boundary effects.

This method can also be adapted to the LHZ layout representing an all-to-all connected graph \cite{Lechner2015}, since it only contains triangluar plaquettes at the chip boundary. 

%

\end{document}